\documentclass[journal,onecolumn,10pt]{WSPLC_class}

\usepackage[left=2cm,top=1.78cm,right=2cm]{geometry}
\usepackage{graphics}
\usepackage{graphicx}
\usepackage{verbatim}
\usepackage{soul}
\usepackage{caption}
\usepackage{subcaption}
\usepackage{epstopdf}
\DeclareGraphicsExtensions{.eps}

\begin{document}

\title{Statistical Assessment of PLC Networking for Front-Hauling in Small Radio Cells}

\author{
	\large
	 Andrea M. Tonello and Francesco Marcuzzi \\\vspace{6pt}
	\normalsize
	 			Alpen-Adria-Universit\"at Klagenfurt, NES, Universitätsstr. 65, 9020 Klagenfurt, Austria, 	andrea.tonello@aau.at\\\vspace{4pt}
}

\markboth{\hspace{0pt}Tenth Workshop on Power Line Communications 10-11 October, 2016, Paris, France}{}


\maketitle

\vspace{-10pt}
\begin{abstract}
The employment of power lines for communications (PLC) has been theorized almost a century ago; although the physical medium is not meant for data transmission, recent technical developments pushed the capacity boundary much higher than expected, allowing to consider PLC for new applications as the one considered in this contribution, i.e., as a solution for small cell back/front-hauling. In principle, back-hauling for cellular networks via PLC is very attractive given how pervasive the power line infrastructure is. Telecom operators are looking for solutions that can handle the traffic increase which is doublying every year. This calls for improvements of both cellular systems and front-hauling technologies that have to bridge the radio network with the core network. Small cells are currently advocated  as the solution to spatially fragment the network and offer high capacity in densely populated areas; this is also due to the availability of spectrum at high frequency (mmWave links) which inherently offers large bandwidth but forces the cells to become smaller due to the short distance coverage. Therefore, it appears that mobile operators may eventually have to deploy an order of magnitude more (small) cells compared to existing networks having macro cells. This translates in a high density cellular network that poses more challenges for the back-haul.\\
\end{abstract}

\section{Introduction}
Nowadays, the back-haul portion of the network relies either on radio or high speed wireline (ADSL, optic fiber) technologies, which usually have to fulfill high constraints on reliability ($99 \% \div 99.99 \% $); typical capacities of radio links are around some hundred Mbps, while cabled connections can reach an order of magnitude or two above this value. Being the capital and operational expenditures (CAPEX and OPEX) a great factor influencing the back-hauling portion of the network, the already deployed power line infrastructure offers an interesting asset for the operators and its exploitation may significant reduce these costs. \\
Indeed, PLC is attractive \cite{PLC,NT1} but a technical challenge exists since it has to be investigated whether PLC has the potentiality to fulfill the requirements. The main aspects to keep in mind when designing a back-haul network are: a) its capacity shall support the customer requirements (traffic) and take into account the core network characteristics; b) it should not exploit end-user spectrum (if radio based); c) it shall be reliable and enable easy deployment and maintenance.\\
While PLC support a few of these aspects, it is not clear what the network requirement is (especially in the domain of small cells) and consequently if PLC can meet coverage and capacity needs.\\
The work presented in this paper addresses the above issues and poses the base for further studies by proposing a new methodology to assess the requirements and performance of PLC for front/back-hauling. In detail, it is herein proposed to bring together and analyse the world of PLC and of small radio cell networks. To do so, three main aspects have to be considered: a) the radio cell network topology; b) the PLC power line topology; c) the traffic to be back-hauled. \\

\section{Cellular-PLC Network Analytical Model}
\begin{figure}[h]
\begin{center}
	\includegraphics[scale = 0.6]{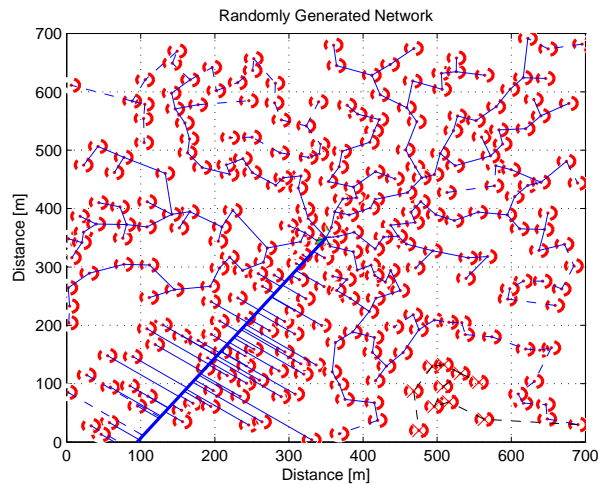}
	\caption{This graph shows an example of a full deployment of a small cell network. All the base stations are topologically interconnected by an underlying system of power lines. Each small cell covers a 400 $m^2$ area, the simulation area considers a squared territory with a 700 m long side and 25\% of its total area covered by small cell services. A total of 6 branches depart from the central concentrator, which is considered the aggregation and access point (hub) towards the core network.}
	\label{fig1}
\end{center}
\end{figure}

We have assessed the three aspects above, i.e. a) the radio cell network topology, b) the PLC power line topology, c) the traffic to be back-hauled, as follows. Firstly, we consider the radio access network that comprises mobile users connected to their base stations (BSs). Base stations are connected to the network via power line cables exploiting the power network that feeds the base stations. More in detail, we consider the front-haul section so that PLC is offered from the BSs to the front-haul hub. Two key issues for the analysis are the BSs displacement and the topology of the power grid connecting them. To obtain the former, i.e., the small cell network layout, we adopt a stochastic geometry approach  \cite{SG1} which is becoming a common approach to evaluate performance of cellular networks. Each small cell is identified by its central radio base station and its area of influence, marked with a dashed red circle in Fig. \ref{fig1}. This model makes overlapping between the coverage area of different cells possible to a certain extent. This is due to the fact that the representation used here accounts for the maximum potential extension of a small cell's coverage, but not for the surrounding environmental conditions: being small cells often employed for indoor coverage, overlaps can account for areas potentially reached by different base stations but actually covered by only one of the nearby stations. Also, being this simulation operated on a two-dimensional plane, overlapping cells can represent cells on different storeys of the same building. The idea of using statistical geometry to assess the performance was used also in the context of in-home PLC networks in \cite{NT2,NT3,NT4}. Herein, a statistically geometrical model for the topology and the wiring structure was proposed. \\

The underneath power grid topology is obtained by building a power delivery network realized according to cost criteria that are followed in building typical distribution grids. The grid is structured as follows: a central concentrator is placed in every MV-LV transformer substation; each transformer feeds a power line branch that can be only a few hundred meters long and can bring power to up to 35 households (according to the European power line paradigm \cite{PLC}). Each one of these branches (referred to as main branches) identifies its geographical sector and it connects via sub branches the radio cells according to a topology. The topologies chosen for the model account for different layouts of the power grid infrastructure; these topologies are:
\begin{itemize}
\item \textbf{Bus Topology}: a central bus is driven through the set of cells in a sector and the radio base stations are connected to it through an orthogonal connection;
\item \textbf{Tree Topology}: cells are interconnected through a branching procedure that links each new cell to the nearest already connected one;
\item \textbf{Chain Topology}: this is actually a different kind of tree topology, since it allows branching to avoid power line intersections, although the chain connection paradigm is preferred.
\end{itemize}
Simulation results show that different topologies have different reachability rates: cells that are deployed far from the central concentrator may bring PLC service to the cells.\\
Eventually, a traffic generation model is developed to simulate realistic voice and data traffic in each cell. While voice calls are simulated through a Poissonian traffic model, data traffic is generated with a long tail paradigm, using data extrapolated from \cite{SCN, LTTM, quora1}. Recent estimates for small cell networks have shown that 97\% of the connection requests are for data, while the remainder accounts for voice and text transmissions. For data transmissions: 80\% of the generated packets have a size smaller than 10 kb, while 10\% of the packets accounts for 90\% of the transmitted information; also while 80\% of transmissions last less than 11 s, 0.1\% of them lasts more than 200 s. For voice connections: the average call lasts about 100 s, while its traffic rate is assumed to be constantly equal to 128 kpbs.\\

\section{Numerical Results}
The performance of the overall hybrid radio/PLC network is assessed via simulation for randomly generated topology layouts, hub position displacements, cell and user densities and traffic intensity. Performance is reported in terms of average supported traffic, average wait time between subsequent connection requests and, most importantly, in terms of number of small cells that can be served, tuned through the density parameter. \\

\begin{figure}[h]
\begin{center}
	\begin{subfigure}[b]{0.45\textwidth}
		\includegraphics[width = \textwidth]{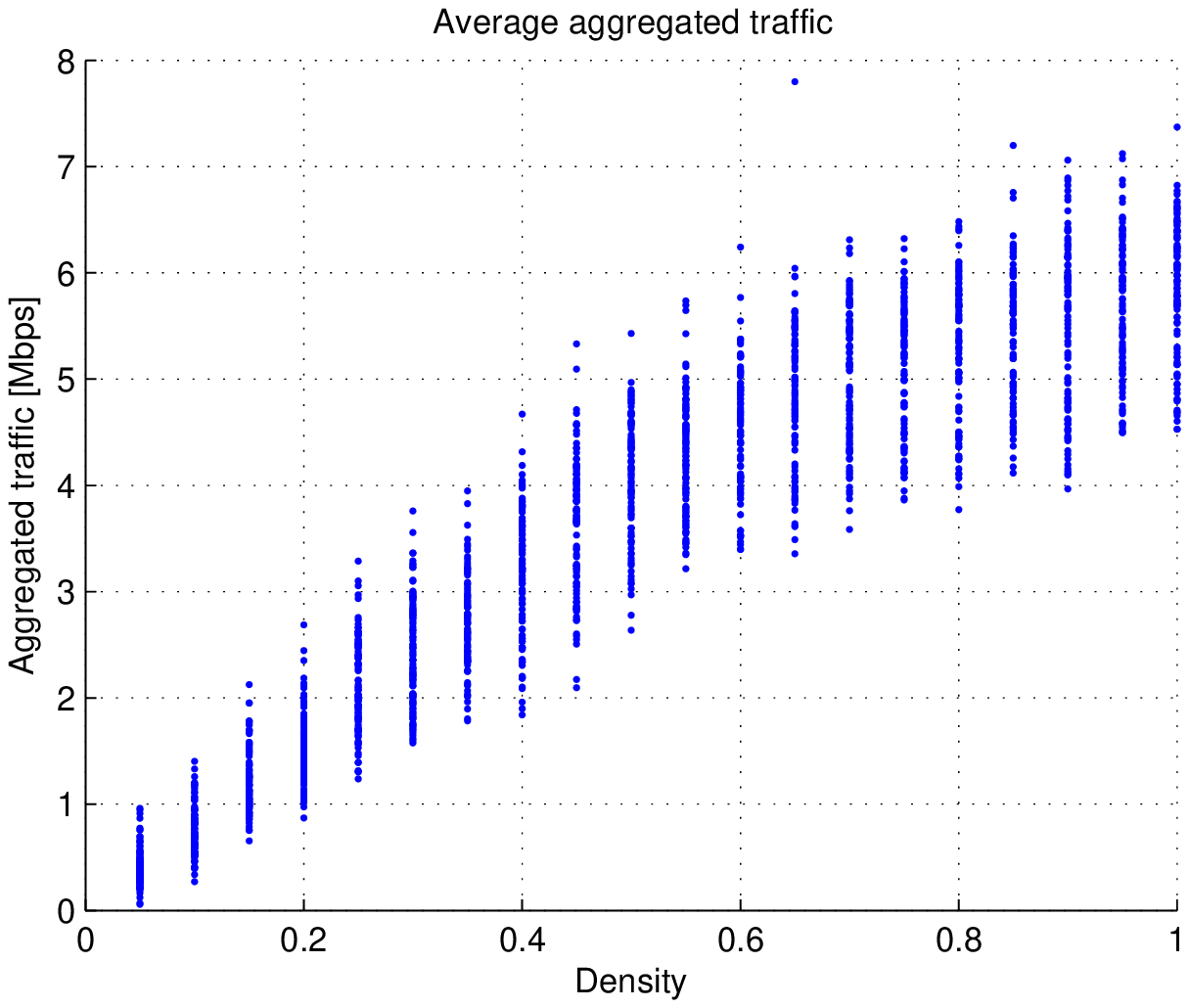}
		\caption{Average Traffic}
		\label{Results1}
	\end{subfigure}
	\begin{subfigure}[b]{0.45\textwidth}
		\includegraphics[width = \textwidth]{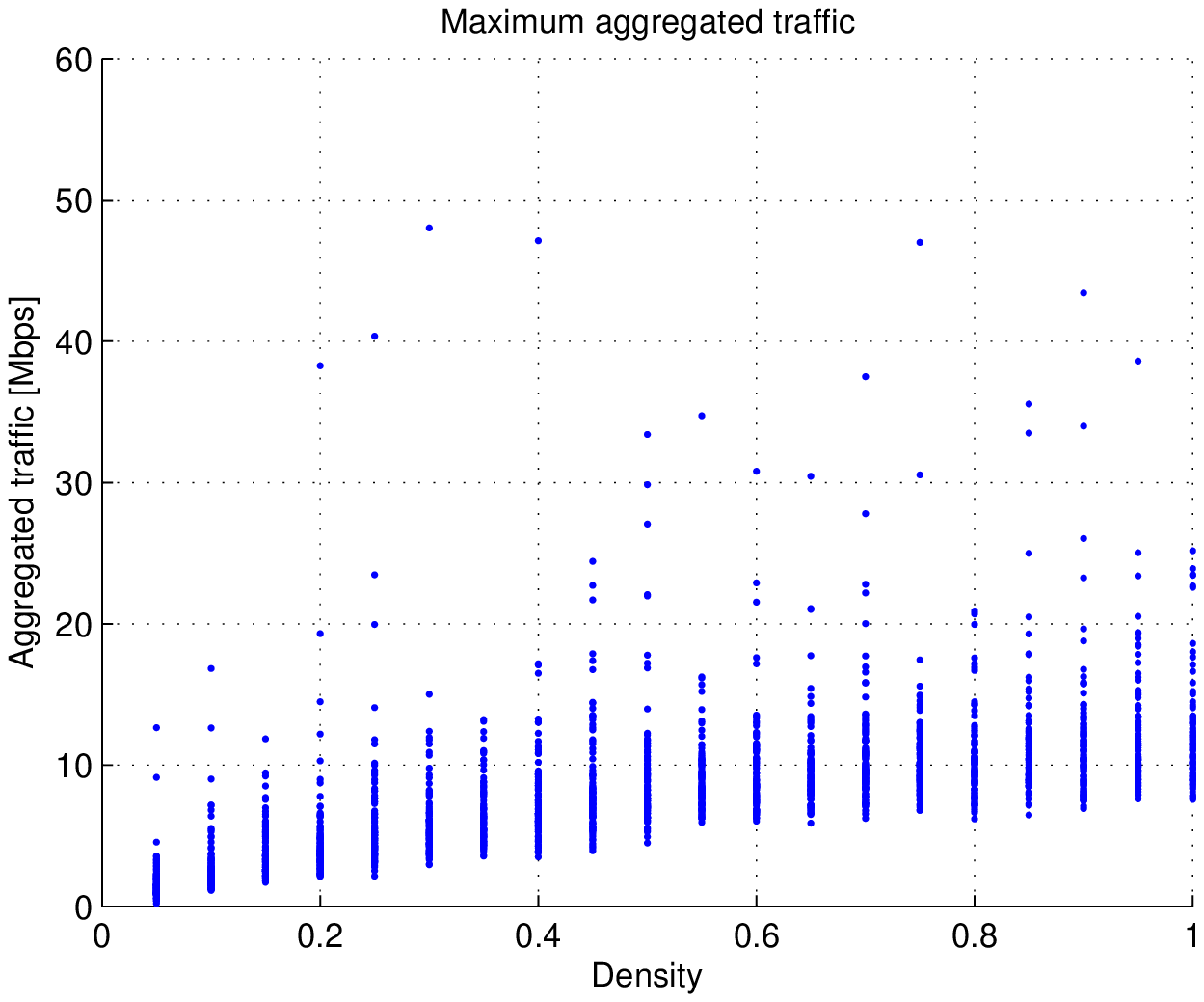}
		\caption{Maximum Traffic}
		\label{Results2}
	\end{subfigure}
	\caption{This figure shows the average and maximum traffic rates in relation to the specified density for the random network generation, where the average time between subsequent calls in each cell is 10 seconds.}
	\label{fig:Results}
\end{center}
\end{figure}

\begin{figure}[ht]
\begin{center}
	\includegraphics[scale = 0.8]{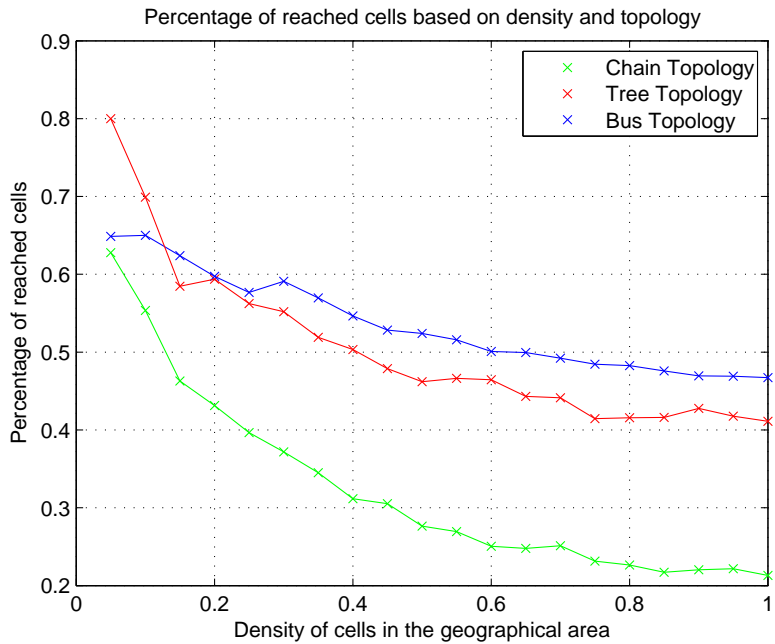}
	\caption{This graph shows the average percentage of reached cells by the PLC service with respect to the density of cells in the geographical area (density = 1 when the geographical area is fully covered). The graph is color coded: blue represents the bus topology, red the tree topology and green the chain one.\\
	If we call M the maximum power line distance a cell can achieve from the central concentrator while still being provided with coverage service and S the side's length of the simulation area, this graph expresses reachability of cells when M = 300 m and S = 700 m.}
	\label{fig:topologies}
\end{center}
\end{figure}

An example is reported in Fig. \ref{fig:Results}, where we show the maximum and average traffic rates in relation to the density of covered geographical territory for a network with a given average time between subsequent calls in each cell (10  seconds). From the analysis, we have found that for the small cell densities that are typical of currently deployed networks and for typical generated mobile traffic, it is required that PLC supports front-haul links in the order of tens of Mbps. The overall performance (sustained network traffic) depends on the way we organize the PLC network and geometrically dispose the hubs that act as data concentrators, as well as it depends on the PLC front-hauling resource management protocol deployed. \\
A numerical analysis of the data from the physical layer revealed how different topologies affect the reachability of far cells; results can be observed in Fig \ref{fig:topologies}. This graph shows how different topologies affect the service coverage and shows that the bus topology has the highest reachability factor, while the implemented chain topology has the lowest.

\section{Conclusions}
From the analysis operated in this work, we have found that for the small cell densities that are typical of currently deployed networks and for typical mobile generated traffic, it is required that PLC supports front-haul links in the order of tens of Mbps, which modern broadband PLC technologies easily support \cite{NT1, NT5}. The overall performance (sustained network traffic) depends on the way we organize the PLC network and geometrically deploy the hubs that act as data concentrators. \\
This study poses the bases for future work in the direction of considering how a PLC front-hauling resource management protocol affects the quality of experience for end users and the requirements for the traffic to be sustained by power line branches; furthermore a routing method together with a resource management protocol can be implemented to increase coverage and overall supported traffic.


\begin{thebibliography}{1}

\bibitem{PLC}
\emph{Power Line Communications: principles, standards, applications from multimedia to smart grid}, L. Lampe, A. M. Tonello, T. Swart (Ed.s), John Wiley \& Sons, 2016.

\bibitem{NT1}
\emph{State-of-the-art in Power Line Communications: from the Applications to the Medium}, C. Cano, A. Pittolo, D. Malone, L. Lampe, A. M. Tonello, A. Dabak, IEEE Journal on Selected Areas in Communications, Vol. 34, Issue 7, pp. 1935 - 1952, July 2016.

\bibitem{NT5}
\emph{Challenges for 1 Gbps Power Line Communications in Home Networks}, A. M. Tonello, P. Siohan, A. Zeddam, X. Mongaboure, Proc. of IEEE 19th Personal Indoor Mobile Radio Communications Symposium (PIMRC) 08, Cannes, France, pp. 1-6, September 15-19, 2008.

\bibitem{SG1}
\emph{Stochastic Geometry for Modeling, Analysis, and Design of Multi-Tier and Cognitive Cellular Wireless Networks: A Survey}, Hesham ElSawy, Ekram Hossain, Martin Haenggi, IEEE Communications Surveys and Tutorials 15(3), July 2013.

\bibitem{NT2}
\emph{Bottom-Up Statistical PLC Channel Modeling – Part I: Random Topology Model and Efficient Transfer Function Computation}, A. M. Tonello, F. Versolatto, IEEE Trans. on Power Delivery, pp. 891-898, April 2011.

\bibitem{NT3}
\emph{Bottom-Up Statistical PLC Channel Modeling – Part II: Inferring the Statistics}, A. M. Tonello, F. Versolatto, Trans. on Power Delivery, pp. 2356-2363, October 2010.
 
\bibitem{NT4}
\emph{A MIMO PLC Random Channel Generator and Capacity Analysis}, F. Versolatto, A. M. Tonello, Proc. ISPLC 2011, Udine, Italy, April 3-6, 2011.

\bibitem{SCN}
\emph{Small Cell Networks}, Tony Q. S. Quek, Guillaume De La Rouche, Ismail Guvenc, Marios Kountouris, Cambridge Press University, 2012.

\bibitem{LTTM}
\emph{Flow-Level Traffic Matrix Generation for Various Data Center Networks}, Yoonseon Han, Sin-seok Seo, Chankyou Hwang, Jae-Hyoung Yoo, James Won-Ki Hong, Network Operations and Management Symposium (NOMS), 2014 IEEE.
 
\bibitem{quora1}
\emph{How long is the average phone call? - Quora}, online discussion: www.quora.com/How-long-is-the-average-phone-call?

\end{thebibliography}
\end{document}